\documentclass[aps,prl,twocolumn,superscriptaddress,showpacs]{revtex4}

\usepackage{graphicx}

\begin{document}


\title{Probing the Kondo density of states in a three-terminal quantum ring}


\author{R. Leturcq}
\email[]{leturcq@phys.ethz.ch}
\author{L. Schmid}
\author{K. Ensslin}
\affiliation{Solid State Physics Laboratory, ETH Z{\"u}rich, 8093 Z{\"u}rich, Switzerland}
\author{Y. Meir}
\affiliation{Department of Physics, Ben Gurion University, Beer Sheva 84105, Israel}
\author{D. C. Driscoll}
\author{A. C. Gossard}
\affiliation{Materials Department, University of California, Santa Barbara, CA 93106, USA}


\date{\today}

\begin{abstract}
We have measured the Kondo effect in a quantum ring connected to three terminals. In this configuration non-linear transport measurements allow to check which lead contributes to the Kondo density of states (DOS) and which does not. When the ring is connected to two strongly and one weakly coupled leads, the measurement of the differential conductance through the weakly coupled lead allows to probe directly the Kondo DOS. In addition, by applying a bias between the two strongly coupled leads, we show the splitting of the out-of-equilibrium DOS.
\end{abstract}

\pacs{72.15.Qm, 73.23.Hk, 75.20.Hr}

\maketitle


The Kondo effect is one of the hallmarks of many-body physics. It was discovered in bulk metals with magnetic impurities providing localized unpaired spins \cite{Kondo01}, and observed later in semiconductor quantum dots \cite{KondoQD} and quantum rings \cite{Keyser01,Fuhrer06}. The electrical tunability of many parameters of quantum dots makes them a favorable system to study the Kondo effect \cite{Kouwenhoven03}, in particular out-of-equilibrium, which is an important theoretical issue \cite{Hershfield02,MeirWingreen,Konig01,Kaminskietal,Rosch01,Kondononeq,Paaske01}. Recently, theoretical proposals have shown that quantum dots might be suitable systems for measuring the local density of states (DOS) in the Kondo regime in and out-of-equilibrium \cite{Sun01,Lebanon01}, a challenge in scanning tunneling experiments \cite{LiJMadhavan}.

\begin{figure}
\includegraphics[width=3.0in]{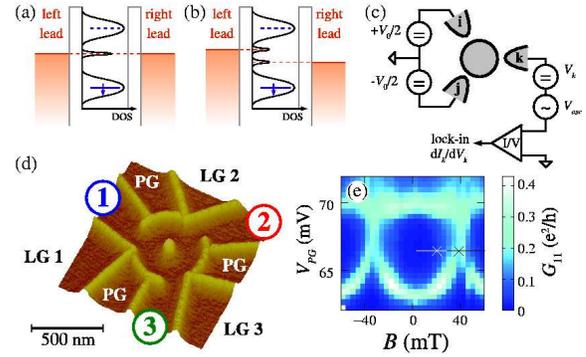}
\caption{Top: Scheme of the DOS in a quantum dot connected to two leads in the Kondo regime (a) in equilibrium and (b) out-of-equilibrium. (c) Experimental set-up for the non-linear transport measurements in the three-terminal configuration. (d) AFM image of the oxide lines defining the three-terminal quantum ring. (e) Linear conductance through terminal 1, $G_{11}$, measured as a function of the magnetic field applied perpendicular to the plane of the 2DEG, $B$, and the plunger gate voltage $V_{PG}$. The horizontal line corresponds to $V_{PG} = 66.1$ mV, and the two crosses to $B = 20$ and 40 mT. \label{fig1}}
\end{figure}

For a quantum dot at equilibrium connected to two leads (see Fig.~\ref{fig1}(a)), the screening of the local spin by electrons from both leads gives rise to a peak in the DOS aligned with the chemical potential of the leads, and with a width of order $k_BT_K$, $T_K$ being the Kondo temperature. When a bias larger than $k_BT_K/e$ is applied between the leads (see Fig.~\ref{fig1}(b)), a splitting of the enhanced DOS into two peaks aligned with the two chemical potentials of the leads has been predicted theoretically \cite{MeirWingreen,Konig01,Rosch01}. These peaks will be reduced relative to the equilibrium peak, due to incoherent scattering between the two leads.

In a two-terminal experiment, the current probes the total DOS between the two chemical potentials. In addition, cotunneling processes involving a spin flip gives rise to a finite conductance value between Coulomb resonances if the dot is the Kondo regime. Thus the differential conductance will be maximal at zero voltage and be suppressed at finite bias (Fig.~\ref{fig1}(b)), giving rise to a peak of the conductance around zero bias, the so-called zero bias anomaly (ZBA) \cite{MeirWingreen,KondoQD}. It is thus not possible to observe the splitting of the DOS in a two-terminal experiment. It has been theoretically proposed that, by adding a third weakly coupled lead, and measuring the conductance through this additional lead, one should be able to measure directly the splitting of the DOS \cite{Sun01,Lebanon01}. A similar splitting of the ZBA has already been observed in a dot connected to one lead and one narrow wire driven out-of-equilibrium \cite{Franceschi01}. However, to probe directly the DOS, one needs to be able to control the coupling of each lead to the dot, and thus to check whether the probing lead is weakly coupled and does not modify the DOS.

\begin{figure*}
\includegraphics[width=5.6in]{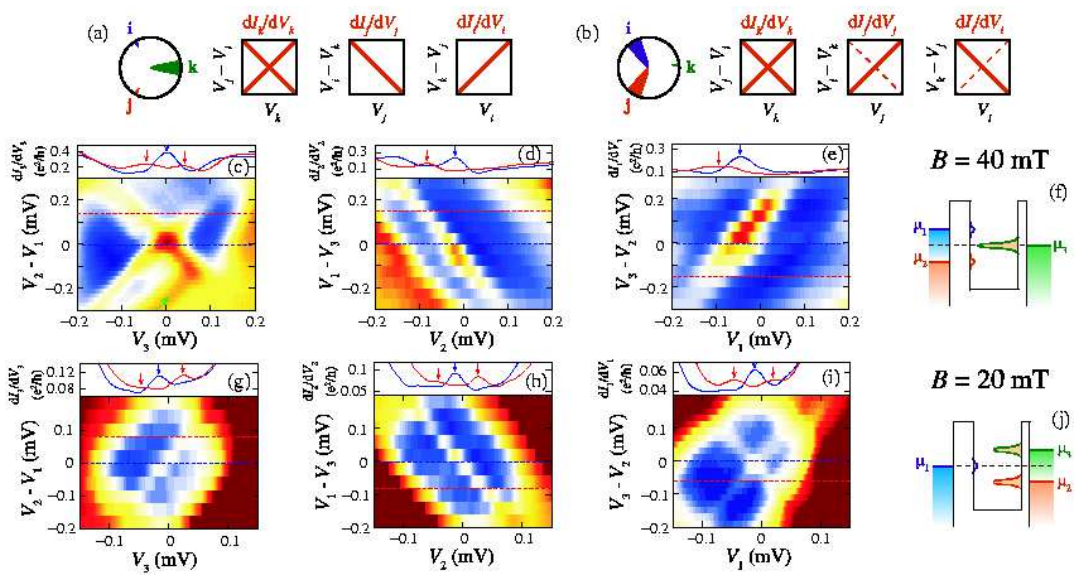}
\caption{Top: Position of the Kondo resonances expected from the measurement with the set-up described in Fig.~\ref{fig1}(c) in the plane ($V_k$, $V_i-V_j$) in the two relevant configurations: one strongly and two weakly coupled leads (a), and two strongly and one weakly coupled leads (b). Straight lines are the expected stronger resonance, while dashed lines are weaker resonances. (c-e) Measurements at $B = 40$ mT. (g-i) Measurements at $B = 20$ mT. The probing lead is number 3 (c,g), 2 (d,h) and 1 (e,i). The color scales are arbitrary, ranging from dark blue (minimum value) to dark red (maximum value). The green arrow in (c) points to a line attributed to an additional resonance in one quantum point contact. The upper curves are cuts of the 2D plots corresponding to the dashed lines. The arrows point on the resonances. (f,j) Schemes of the configurations probed at $B = 40$ mT (f) and at $B = 20$ mT (j). The magnitude of the peaks represents the strength of the coupling to each lead. \label{fig2}}
\end{figure*}


Here we report a novel geometry -- a three terminal quantum ring -- that allows unambiguous determination of which leads are strongly coupled and which are weakly coupled to the ring. We show two configurations, the single strongly coupled lead regime (and two weakly coupled ones) and the single weakly coupled lead regime. The later regime allows us to probe the DOS in the Kondo regime in and out-of-equilibrium. In addition, we show that the probing lead does not significantly modify the enhanced DOS due to the Kondo effect, which is the condition to probe directly the DOS.


The sample was fabricated on an AlGaAs-GaAs heterostructure containing a two-dimensional electron gas (2DEG) 34 nm below the surface (electron density $4.5 \times 10^{15}$ m$^{-2}$, mobility 25 m$^2$(Vs)$^{-1}$). An atomic force microscope (AFM) was used to locally oxydize the surface of the semiconductor \cite{HeldFuhrer}. The 2DEG is depleted below the oxide lines, defining the three-terminal quantum ring shown in Fig.~\ref{fig1}(d). The three leads labelled 1 to 3 are coupled to the ring through three quantum point contacts controlled by the lateral gates LG1 to LG3. The number of electrons in the ring is controlled by the three plunger gates PG set at the same potential $V_{PG}$. For most measurements, $V_{PG}$ is fixed to 66.1 mV. The measurements have been done in a 3He/4He dilution refrigerator with an electronic temperature less than 50 mK. The measured Aharonov-Bohm period is 75 mT, corresponding to an approximate diameter of the ring of 270 nm. Measurements of Coulomb diamonds in the weak-coupling regime reveal a charging energy $E_C \approx 1.2$ meV and an average single-particle level spacing $\Delta E$ $\approx$ 200 $\mu$eV.


The coupling to each lead can be controlled either by electrostatic gates, or by changing the magnetic field, which changes the distribution of the wave function in the dot \cite{Leturcq04}. Here we keep the gate voltages constant and change the magnetic field $B$, allowing a finer tuning of the couplings. Figure~\ref{fig1}(e) shows the linear conductance matrix element $G_{11} = I_1/V_1$ as a function of $V_{PG}$ and $B$, where the current through lead 1, $I_1$, is measured when applying a quasi-DC bias $V_1 = \pm 10$ $\mu$V on lead 1, leads 2 and 3 being grounded (more details are given in Ref.~\onlinecite{Leturcq04}). At zero magnetic field we see two conductance peaks around $V_{PG} = 63$ and 70 mV corresponding to the usual conductance resonances. The positions of both peaks oscillate as a function of the magnetic field, as already demonstrated for extended states in a quantum ring \cite{Fuhr01}, and fluctuations of their magnitudes are related to fluctuations of the couplings.

Clear characteristics of the Kondo effect are seen in the valley inbetween the two peaks. A ZBA is measured at $V_{PG} = 66.1$ mV and both $B = 40$ mT (see Fig.~\ref{fig2}(c)) and $B = 20$ mT (see Fig.~\ref{fig2}(i)). In addition we have checked that the ZBA is suppressed when increasing the temperature, and stays at zero bias when changing $V_{PG}$. From the temperature dependence of the maximum of the ZBA we have extracted a Kondo temperature \cite{GoldhaberVdWiel} of $T_K = 500 \pm 50$ mK at $B = 40$ mT and $T_K \approx 150$ to 200 mK at $B = 20$ mT. Modulation of the Kondo temperature with a perpendicular magnetic field has already been observed in similar quantum rings \cite{Fuhrer06} and explained via modulation of the wave function in the ring due to the Aharonov-Bohm effect. 

For a three-terminal quantum dot one can distinguish eight different configurations, for which each lead can be either strongly coupled to the dot, giving rise to a Kondo resonance, or weakly coupled leaving the DOS basically unaffected. Here we demonstrate an experimental method to distinguish between these configurations. The non-linear transport measurements are made with the set-up described in Fig.~\ref{fig1}(c). A DC bias is applied to the leads $i$ and $j$ ($V_i-V_j = V_0$), while the differential conductance $dI_k/dV_k$ is measured with a lock-in technique ($V_{osc} = 2$ $\mu$V, frequency 13.73 Hz) through the probing lead $k$ as a function of a DC bias $V_k$. By rotating the set-up, we obtain three measurements corresponding to $k = 3$, 2 and 1. We now follow the Kondo peak in the plane ($V_k$, $V_i-V_j$), for each $k$. If all leads are weakly coupled, no peak will be observed. If all leads are strongly coupled a cross pattern will be observed in all three measurements, which means that there will be a Kondo peak whenever two chemical potentials cross. In the other two cases of one or two strongly coupled leads, sketched in Figs.~\ref{fig2}(a) and (b), one expects one cross pattern and two diagonal patterns: the cross pattern arises when, as one scans the chemical potential in lead $k$, a Kondo resonance arises as it crosses the chemical potentials in both lead $i$ and lead $j$. Since, in order to observe a Kondo effect upon crossing, only one of the respective leads needs to be strongly coupled, this cross pattern can arise when either only lead $k$ is strongly coupled, or when both lead $i$ and lead $j$ are strongly coupled and lead $k$ is not. These two cases can be distinguished by scanning the voltage in lead $j$ (resp. $i$) and checking whether a Kondo peak arises when it crosses lead $k$ (first case, one strongly coupled lead -- $k$ -- in Fig.~\ref{fig2}(a)) or lead $i$ (resp. $j$) (second case, two strongly coupled leads -- $i$ and $j$ -- in Fig.~\ref{fig2}(b)). This feature of the set-up is the key to be able to probe the DOS in a Kondo system. Such a procedure is not straightforward in a quantum dot connected to two-terminals \cite{Franceschi01}.

A comparison between the expected result (Figs.~\ref{fig2}(a) and (b)) and the measurements at $B = 40$ mT  (Figs.~\ref{fig2}(c) to (e)) shows that in this case the ring is strongly coupled to one lead (number 3), the two others being weakly coupled, as sketched in Fig.~\ref{fig2}(f). A similar analysis for the case of $B = 20$ mT  (Figs.~\ref{fig2}(g) to (i)) shows that the ring is strongly coupled to two leads, and lead 1 is weakly coupled, as sketched in Fig.~\ref{fig2}(j). In the following we focus on this configuration at $B = 20$ mT, which is the proper one to probe the out-of-equilibrium Kondo DOS.

\begin{figure}
\includegraphics[width=3.4in]{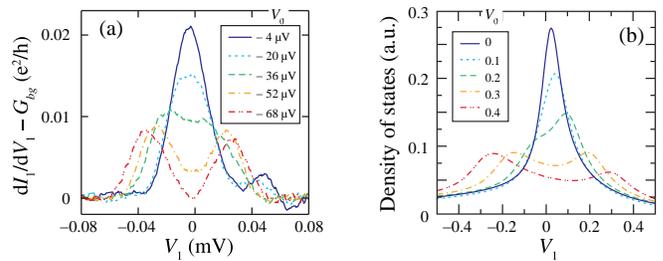}
\caption{(a) Difference between the differential conductance $dI_1/dV_1$ (from Fig.~\ref{fig2}(i)) and the background conductance $G_{bg}$ at $B = 20$ mT. With $k_BT_K \approx 15$ $\mu$eV, the bias voltages $V_0 = V_3 - V_2$ correspond approximately to 0, 1, 2, 3 and 4 times $k_BT_K/e$. (b) Calculation of the enhanced DOS vs. energy $V_1$ for different bias voltages $V_0$ applied between two leads corresponding to the same ratios as for the experimental curves. The couplings to the leads are respectively 0.4 and 0.6, the bare energy (in this case, of a hole) is $-2$, the temperature is 0.03 and the Kondo temperature 0.05. A Lorentzian fit of the bare energy peak has been subtracted from the numerical curve. All energies are in terms of the level broadening. \label{fig3}}
\end{figure}

The results corresponding to probing the out-of-equilibrium Kondo DOS at $B = 20$ mT are shown in Fig.~\ref{fig2}(i). In Fig.~\ref{fig3}(a), part of these data are shown after removing the background conductance $G_{bg}$ in order to emphasize the Kondo resonance. $G_{bg}$ is deduced from a fit by the sum of a Lorentzian and a linear function, corresponding to the bare single-particle levels. When increasing the bias voltage $V_0=V_3-V_2$, the magnitude of the peak first decreases by a factor of about 2. Above $V_0 \approx 2 \times k_BT_K/e$, the Kondo peak splits into two peaks. The two peaks vanish slowly when increasing the bias voltage further, and are not resolved anymore above $V_0 \approx 10 - 20 \times k_BT_K/e$. In Fig.~\ref{fig3}(b), the calculated out-of-equilibrium Kondo DOS is shown for several bias voltages. The calculation is made for the infinite-interaction single-impurity Anderson model, using the noncrossing-approximation \cite{Bickers01} generalized to nonequilibrium \cite{MeirWingreen}. We see a good agreement between the experiment and the theory. To best fit the data, unequal coupling to the two leads (ratio of 2:3) was assumed in the numerical calculation. Unequal couplings are, of course, expected experimentally. The two peaks observed in the differential conductance can then be attributed to the splitting of the Kondo DOS for $V_0 > 2 \times k_BT_K/e$ \cite{MeirWingreen,Konig01,Rosch01}. At higher voltages, the further decrease of the peaks in the experiment is attributed to the dephasing of the Kondo state by the finite bias \cite{MeirWingreen,Kaminskietal,Paaske01}.

\begin{figure}
\includegraphics[width=3.4in]{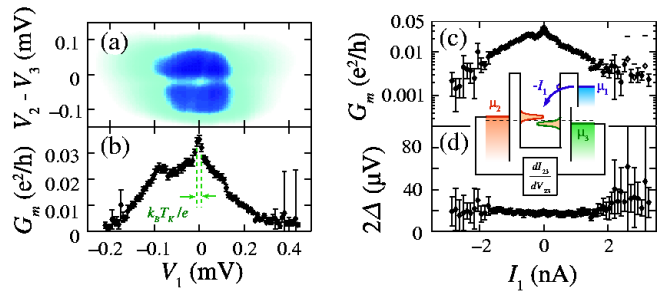}
\caption{(a) $dI_{23}/dV_{23}$ measured as a function of $V_{23}$ and $V_1$. The color scale ranges from 0.05 $e^2/h$ (dark blue) to 0.2 $e^2/h$ (white). (b-d) Magnitude $G_m$ and full width at half maximum $2 \Delta$ of the peak in $dI_{23}/dV_{23}$ vs. $V_{23}$ as a function of $V_1$ (b) and $I_1$ (c,d). The error bars correspond to the 95 \% confidence bounds of the fit of the peaks. Inset: Scheme of the measurement for testing the influence of the probing lead. \label{fig4}}
\end{figure}

While it was argued theoretically that an experimental estimate of the out-of-equilibrium DOS can be achieved even if the probing lead is as strongly coupled as the two others \cite{Lebanon01,ChoSY01}, we have good reason to believe that in our case, it is indeed significantly more weakly coupled. The first argument is shown in Figs.~\ref{fig2}(g) and (h), where a strong peak arises when aligning the chemical potentials of leads 2 and 3, and a very weak peak is observed when aligning the chemical potential of lead 1 with either 2 or 3 (see red arrows in upper figures). A second check is to study the effect of lead 1 on the usual two-terminal non-linear transport through the two strongly coupled leads 2 and 3, as sketched in the inset of Fig.~\ref{fig4}. The differential conductance between leads 2 and 3, $dI_{23}/dV_{23}$, is measured with an AC technique as a function of a DC bias $V_{23} = V_2-V_3$ symmetrically applied between leads 2 and 3. In addition a DC bias $V_1$ is applied on lead 1 and the DC current flowing from lead 1, $I_1$, is measured. $dI_{23}/dV_{23}$ vs. $V_{23}$ shows a ZBA in a large range of $V_1$ (see Fig.~\ref{fig4}(a)). To evaluate the evolution of the ZBA as a function of $V_1$ we have fitted $dI_{23}/dV_{23}$ vs. $V_{23}$ with a linear background and a Lorentzian of magnitude $G_m$ and full width at half maximum $2 \Delta$. Figure~\ref{fig4}(b) shows that the ZBA is observed for voltages as high as $V_1 > 10 \times k_BT_K/e$, and only a weak decrease of the peak amplitude is observed for $V_1 \approx k_BT_K/e$. We conclude that the perturbation of the Kondo DOS created by leads 2 and 3 by the weakly coupled lead 1 is almost negligible, and that the curves in Fig.~\ref{fig3}(a) are very close to the out-of-equilibrium Kondo DOS convoluted by the Fermi distribution in lead 1 \cite{Lebanon01}.

When increasing $V_1$ further, Fig.~\ref{fig4}(b) shows a continuous decrease of the amplitude of the peak. As a function of the current through lead 1, this decrease is almost exponential (see Fig.~\ref{fig4}(c)). The Kondo resonance is expected to be suppressed by lead 1 due to dephasing processes, such as the exchange of electrons between lead 1 and the ring, thus destroying the coherent state in the ring \cite{Sanchez01}, or the shot noise due to the current flowing through lead 1 \cite{Silva01,Avinun01}. However, dephasing is also expected to lead to an increase of the peak width \cite{MeirWingreen,Kaminskietal,Paaske01}. Figure~\ref{fig4}(d) shows that no significant change of the peak width is seen within the error bars up to voltages $V_1 > 5 \times k_BT_K/e$. This indicates that the process leading to the decrease of the peak amplitude in our experiment is most likely not related to dephasing. A scenario leading to a decrease of only the Kondo peak height has been proposed for a quantum dot in the Kondo regime in weak interaction with a quantum point contact \cite{Silva01}. The nature of this process, and whether it can be applied to our case, is an open question.


Our experiments demonstrate that a three-terminal set-up allows to determine the contribution of each lead to the Kondo DOS in a quantum dot. This allows us to measure the Kondo DOS in and out-of-equilibrium. Compared to the generally used two-terminal tunneling measurements, the multi-terminal configuration gives additional insight into the energy spectrum, wave function coupling and many-particle DOS of a quantum system.

\begin{acknowledgments}
The authors thank T. Ihn, K. Kobayashi, S. De Franceschi, L. P. Kouwenhoven and  A. Rosch for usefull discussions. Financial support from the Swiss Science Foundation (Schweizerischer Nationalfonds) and from the EU Human Potential Program financed via the Bundesministerium f{\"u}r Bildung und Wissenschaft is gratefully acknowledged.
\end{acknowledgments}

\bibliography{biblio}

\begin{thebibliography}{35}
\expandafter\ifx\csname natexlab\endcsname\relax\def\natexlab#1{#1}\fi
\expandafter\ifx\csname bibnamefont\endcsname\relax
  \def\bibnamefont#1{#1}\fi
\expandafter\ifx\csname bibfnamefont\endcsname\relax
  \def\bibfnamefont#1{#1}\fi
\expandafter\ifx\csname citenamefont\endcsname\relax
  \def\citenamefont#1{#1}\fi
\expandafter\ifx\csname url\endcsname\relax
  \def\url#1{\texttt{#1}}\fi
\expandafter\ifx\csname urlprefix\endcsname\relax\def\urlprefix{URL }\fi
\providecommand{\bibinfo}[2]{#2}
\providecommand{\eprint}[2][]{\url{#2}}

\bibitem[{\citenamefont{Kondo}(1964)}]{Kondo01}
\bibinfo{author}{\bibfnamefont{J.}~\bibnamefont{Kondo}},
  \bibinfo{journal}{Prog. Theor. Phys.} \textbf{\bibinfo{volume}{32}},
  \bibinfo{pages}{37} (\bibinfo{year}{1964}).

\bibitem{KondoQD}
\bibinfo{author}{\bibfnamefont{D.}~\bibnamefont{Goldhaber-Gordon}}
  \bibinfo{author}{{\it et al.}},
  \bibinfo{journal}{Nature} \textbf{\bibinfo{volume}{391}},
  \bibinfo{pages}{156} (\bibinfo{year}{1998}{\natexlab{a}});
\bibinfo{author}{\bibfnamefont{S.~M.} \bibnamefont{Cronenwett}},
  \bibinfo{author}{\bibfnamefont{T.~H.} \bibnamefont{Oosterkamp}},
  \bibnamefont{and} \bibinfo{author}{\bibfnamefont{L.~P.}
  \bibnamefont{Kouwenhoven}}, \bibinfo{journal}{Science}
  \textbf{\bibinfo{volume}{281}}, \bibinfo{pages}{540} (\bibinfo{year}{1998});
\bibinfo{author}{\bibfnamefont{J.}~\bibnamefont{Schmid}},
  \bibinfo{author}{\bibfnamefont{J.}~\bibnamefont{Weis}},
  \bibinfo{author}{\bibfnamefont{K.}~\bibnamefont{Eberl}}, \bibnamefont{and}
  \bibinfo{author}{\bibfnamefont{K.}~\bibnamefont{v.~Klitzing}},
  \bibinfo{journal}{Physica B} \textbf{\bibinfo{volume}{256-258}},
  \bibinfo{pages}{182} (\bibinfo{year}{1998}).

\bibitem[{\citenamefont{Keyser et~al.}(2003)\citenamefont{Keyser, F{\"u}hner,
  Borck, Haug, Bichler, Abstreiter, and Wegscheider}}]{Keyser01}
\bibinfo{author}{\bibfnamefont{U.~F.} \bibnamefont{Keyser}}
  \bibinfo{author}{{\it et al.}},
  \bibinfo{journal}{Phys. Rev. Lett.} \textbf{\bibinfo{volume}{90}},
  \bibinfo{pages}{196601} (\bibinfo{year}{2003}).

\bibitem[{\citenamefont{Fuhrer et~al.}(2004)\citenamefont{Fuhrer, Ihn, Ensslin,
  Wegscheider, and Bichler}}]{Fuhrer06}
\bibinfo{author}{\bibfnamefont{A.}~\bibnamefont{Fuhrer}},
\bibinfo{author}{\bibfnamefont{T.}~\bibnamefont{Ihn}},
\bibinfo{author}{\bibfnamefont{K.}~\bibnamefont{Ensslin}},
\bibinfo{author}{\bibfnamefont{W.}~\bibnamefont{Wegscheider}} \bibnamefont{and}
\bibinfo{author}{\bibfnamefont{M.}~\bibnamefont{Bichler}},
  \bibinfo{journal}{Phys. Rev. Lett.} \textbf{\bibinfo{volume}{93}},
  \bibinfo{pages}{176803} (\bibinfo{year}{2004}).

\bibitem[{\citenamefont{Kouwenhoven and Glazman}(2001)}]{Kouwenhoven03}
\bibinfo{author}{\bibfnamefont{L.}~\bibnamefont{Kouwenhoven}} \bibnamefont{and}
  \bibinfo{author}{\bibfnamefont{L.}~\bibnamefont{Glazman}},
  \bibinfo{journal}{Physics World} \textbf{\bibinfo{volume}{14}},
  \bibinfo{pages}{33} (\bibinfo{year}{2001}).

\bibitem[{\citenamefont{Hershfield et~al.}(1992)\citenamefont{Hershfield,
  Davies, and Wilkins}}]{Hershfield02}
\bibinfo{author}{\bibfnamefont{S.}~\bibnamefont{Hershfield}},
  \bibinfo{author}{\bibfnamefont{J.~H.} \bibnamefont{Davies}},
  \bibnamefont{and} \bibinfo{author}{\bibfnamefont{J.~W.}
  \bibnamefont{Wilkins}}, \bibinfo{journal}{Phys. Rev. B}
  \textbf{\bibinfo{volume}{46}}, \bibinfo{pages}{7046} (\bibinfo{year}{1992}).

\bibitem{MeirWingreen}
\bibinfo{author}{\bibfnamefont{Y.}~\bibnamefont{Meir}},
  \bibinfo{author}{\bibfnamefont{N.~S.} \bibnamefont{Wingreen}},
  \bibnamefont{and} \bibinfo{author}{\bibfnamefont{P.~A.} \bibnamefont{Lee}},
  \bibinfo{journal}{Phys. Rev. Lett.} \textbf{\bibinfo{volume}{70}},
  \bibinfo{pages}{2601} (\bibinfo{year}{1993});
\bibinfo{author}{\bibfnamefont{N.~S.} \bibnamefont{Wingreen}} \bibnamefont{and}
  \bibinfo{author}{\bibfnamefont{Y.}~\bibnamefont{Meir}},
  \bibinfo{journal}{Phys. Rev. B} \textbf{\bibinfo{volume}{49}},
  \bibinfo{pages}{11040} (\bibinfo{year}{1994}).

\bibitem[{\citenamefont{K\"onig et~al.}(1996)\citenamefont{K\"onig, Schmid, and
  Schoeller}}]{Konig01}
\bibinfo{author}{\bibfnamefont{J.}~\bibnamefont{K\"onig}},
  \bibinfo{author}{\bibfnamefont{J.}~\bibnamefont{Schmid}},
  \bibinfo{author}{\bibfnamefont{H.}~\bibnamefont{Schoeller}} \bibnamefont{and}
  \bibinfo{author}{\bibfnamefont{G.}~\bibnamefont{Sch\"on}},
  \bibinfo{journal}{Phys. Rev. B} \textbf{\bibinfo{volume}{54}},
  \bibinfo{pages}{16820} (\bibinfo{year}{1996}).

\bibitem{Kaminskietal}
\bibinfo{author}{\bibfnamefont{A.}~\bibnamefont{Kaminski}},
  \bibinfo{author}{\bibfnamefont{Y.~V.} \bibnamefont{Nazarov}},
  \bibnamefont{and} \bibinfo{author}{\bibfnamefont{L.~I.}
  \bibnamefont{Glazman}}, \bibinfo{journal}{Phys. Rev. Lett.}
  \textbf{\bibinfo{volume}{83}}, \bibinfo{pages}{384} (\bibinfo{year}{1999});
\bibinfo{journal}{Phys. Rev. B}
  \textbf{\bibinfo{volume}{62}}, \bibinfo{pages}{8154} (\bibinfo{year}{2000}).

\bibitem[{\citenamefont{Rosch et~al.}(2001)\citenamefont{Rosch, Kroha, and
  W\"olfle}}]{Rosch01}
\bibinfo{author}{\bibfnamefont{A.}~\bibnamefont{Rosch}},
  \bibinfo{author}{\bibfnamefont{J.}~\bibnamefont{Kroha}}, \bibnamefont{and}
  \bibinfo{author}{\bibfnamefont{P.}~\bibnamefont{W\"olfle}},
  \bibinfo{journal}{Phys. Rev. Lett.} \textbf{\bibinfo{volume}{87}},
  \bibinfo{pages}{156802} (\bibinfo{year}{2001}).

\bibitem{Kondononeq}
\bibinfo{author}{\bibfnamefont{P.}~\bibnamefont{Coleman}},
  \bibinfo{author}{\bibfnamefont{C.}~\bibnamefont{Hooley}}, \bibnamefont{and}
  \bibinfo{author}{\bibfnamefont{O.}~\bibnamefont{Parcollet}},
  \bibinfo{journal}{Phys. Rev. Lett.} \textbf{\bibinfo{volume}{86}},
  \bibinfo{pages}{4088} (\bibinfo{year}{2001});
\bibinfo{author}{\bibfnamefont{Y.-W.} \bibnamefont{Lee}} \bibnamefont{and}
  \bibinfo{author}{\bibfnamefont{Y.-L.} \bibnamefont{Lee}},
  \bibinfo{journal}{Phys. Rev. B} \textbf{\bibinfo{volume}{65}},
  \bibinfo{pages}{155324} (\bibinfo{year}{2002});
\bibinfo{author}{\bibfnamefont{T.}~\bibnamefont{Fujii}} \bibnamefont{and}
  \bibinfo{author}{\bibfnamefont{K.}~\bibnamefont{Ueda}},
  \bibinfo{journal}{Phys. Rev. B} \textbf{\bibinfo{volume}{68}},
  \bibinfo{pages}{155310} (\bibinfo{year}{2003});
\bibinfo{author}{\bibfnamefont{J.}~\bibnamefont{Paaske}},
  \bibinfo{author}{\bibfnamefont{A.}~\bibnamefont{Rosch}}, \bibnamefont{and}
  \bibinfo{author}{\bibfnamefont{P.}~\bibnamefont{W\"olfle}},
  \bibinfo{journal}{Phys. Rev. B} \textbf{\bibinfo{volume}{69}},
  \bibinfo{pages}{155330} (\bibinfo{year}{2004}{\natexlab{a}}).

\bibitem[{\citenamefont{Paaske et~al.}(2004{\natexlab{b}})\citenamefont{Paaske,
  Rosch, Kroha, and W\"olfle}}]{Paaske01}
\bibinfo{author}{\bibfnamefont{J.}~\bibnamefont{Paaske}},
  \bibinfo{author}{\bibfnamefont{A.}~\bibnamefont{Rosch}},
  \bibinfo{author}{\bibfnamefont{J.}~\bibnamefont{Kroha}}, \bibnamefont{and}
  \bibinfo{author}{\bibfnamefont{P.}~\bibnamefont{W\"olfle}},
  \bibinfo{journal}{Phys. Rev. B} \textbf{\bibinfo{volume}{70}},
  \bibinfo{pages}{155301} (\bibinfo{year}{2004}{\natexlab{b}}).

\bibitem[{\citenamefont{Sun and Guo}(2001)}]{Sun01}
\bibinfo{author}{\bibfnamefont{Q.-F.}~\bibnamefont{Sun}} \bibnamefont{and}
  \bibinfo{author}{\bibfnamefont{H.}~\bibnamefont{Guo}},
  \bibinfo{journal}{Phys. Rev. B} \textbf{\bibinfo{volume}{64}},
  \bibinfo{pages}{153306} (\bibinfo{year}{2001}).

\bibitem[{\citenamefont{Lebanon and Schiller}(2001)}]{Lebanon01}
\bibinfo{author}{\bibfnamefont{E.}~\bibnamefont{Lebanon}} \bibnamefont{and}
  \bibinfo{author}{\bibfnamefont{A.}~\bibnamefont{Schiller}},
  \bibinfo{journal}{Phys. Rev. B} \textbf{\bibinfo{volume}{65}},
  \bibinfo{pages}{035308} (\bibinfo{year}{2001}).

\bibitem[{\citenamefont{Li et~al.}(1998)\citenamefont{Li, Schneider, Berndt,
  and Delley}}]{LiJMadhavan}
\bibinfo{author}{\bibfnamefont{J.}~\bibnamefont{Li}},
  \bibinfo{author}{\bibfnamefont{W.-D.} \bibnamefont{Schneider}},
  \bibinfo{author}{\bibfnamefont{R.}~\bibnamefont{Berndt}}, \bibnamefont{and}
  \bibinfo{author}{\bibfnamefont{B.}~\bibnamefont{Delley}},
  \bibinfo{journal}{Phys. Rev. Lett.} \textbf{\bibinfo{volume}{80}},
  \bibinfo{pages}{2893} (\bibinfo{year}{1998});
\bibinfo{author}{\bibfnamefont{V.}~\bibnamefont{Madhavan}},
  \bibinfo{author}{{\it et al.}},
  \bibinfo{journal}{Science}
  \textbf{\bibinfo{volume}{280}}, \bibinfo{pages}{567} (\bibinfo{year}{1998}).

\bibitem[{\citenamefont{De~Franceschi et~al.}(2002)\citenamefont{De~Franceschi,
  Hanson, van~der Wiel, Elzerman, Wijpkema, Fujisawa, Tarucha, and
  Kouwenhoven}}]{Franceschi01}
\bibinfo{author}{\bibfnamefont{S.}~\bibnamefont{De~Franceschi}}
  \bibinfo{author}{{\it et al.}},
  \bibinfo{journal}{Phys. Rev. Lett.} \textbf{\bibinfo{volume}{89}},
  \bibinfo{pages}{156801} (\bibinfo{year}{2002}).

\bibitem{HeldFuhrer}
\bibinfo{author}{\bibfnamefont{R.}~\bibnamefont{Held}}
  \bibinfo{author}{{\it et al.}},
  \bibinfo{journal}{Appl. Phys. Lett} \textbf{\bibinfo{volume}{73}},
  \bibinfo{pages}{262} (\bibinfo{year}{1998});
\bibinfo{author}{\bibfnamefont{A.}~\bibnamefont{Fuhrer}}
  \bibinfo{author}{{\it et al.}},
  \bibinfo{journal}{Superl. and Microstruc.} \textbf{\bibinfo{volume}{31}},
  \bibinfo{pages}{19} (\bibinfo{year}{2002}).

\bibitem[{\citenamefont{Leturcq et~al.}(2004)\citenamefont{Leturcq, Graf, Ihn,
  Ensslin, Driscoll, and Gossard}}]{Leturcq04}
\bibinfo{author}{\bibfnamefont{R.}~\bibnamefont{Leturcq}}
  \bibinfo{author}{{\it et al.}},
  \bibinfo{journal}{Europhys. Lett.}
  \textbf{\bibinfo{volume}{67}}, \bibinfo{pages}{439} (\bibinfo{year}{2004}).

\bibitem[{\citenamefont{Fuhrer et~al.}(2001)\citenamefont{Fuhrer, L{\"u}scher,
  Ihn, Heinzel, Ensslin, Wegscheider, and Bichler}}]{Fuhr01}
\bibinfo{author}{\bibfnamefont{A.}~\bibnamefont{Fuhrer}}
  \bibinfo{author}{{\it et al.}},
  \bibinfo{journal}{Nature} \textbf{\bibinfo{volume}{413}},
  \bibinfo{pages}{822} (\bibinfo{year}{2001}).
  
\bibitem{GoldhaberVdWiel}
\bibinfo{author}{\bibfnamefont{D.}~\bibnamefont{Goldhaber-Gordon}}
  \bibinfo{author}{{\it et al.}},
  \bibinfo{journal}{Phys. Rev. Lett.} \textbf{\bibinfo{volume}{81}},
  \bibinfo{pages}{5225} (\bibinfo{year}{1998}{\natexlab{b}});
\bibinfo{author}{\bibfnamefont{W.~G.} \bibnamefont{van~der Wiel}}
  \bibinfo{author}{{\it et al.}},
  \bibinfo{journal}{Science} \textbf{\bibinfo{volume}{289}},
  \bibinfo{pages}{2105} (\bibinfo{year}{2000}).

\bibitem[{\citenamefont{Bickers}(1987)}]{Bickers01}
\bibinfo{author}{\bibfnamefont{N.~E.} \bibnamefont{Bickers}},
  \bibinfo{journal}{Rev. Mod. Phys.} \textbf{\bibinfo{volume}{59}},
  \bibinfo{pages}{845} (\bibinfo{year}{1987}).

\bibitem[{\citenamefont{Cho et~al.}(2003)\citenamefont{Cho, Zhou, and
  McKenzie}}]{ChoSY01}
\bibinfo{author}{\bibfnamefont{S.~Y.} \bibnamefont{Cho}},
  \bibinfo{author}{\bibfnamefont{H.-Q.} \bibnamefont{Zhou}}, \bibnamefont{and}
  \bibinfo{author}{\bibfnamefont{R.~H.} \bibnamefont{McKenzie}},
  \bibinfo{journal}{Phys. Rev. B} \textbf{\bibinfo{volume}{68}},
  \bibinfo{pages}{125327} (\bibinfo{year}{2003}).

\bibitem[{\citenamefont{S\'anchez and L\'opez}(2005)}]{Sanchez01}
\bibinfo{author}{\bibfnamefont{D.}~\bibnamefont{S\'anchez}} \bibnamefont{and}
  \bibinfo{author}{\bibfnamefont{R.}~\bibnamefont{L\'opez}},
  \bibinfo{journal}{Phys. Rev. B} \textbf{\bibinfo{volume}{71}},
  \bibinfo{pages}{035315} (\bibinfo{year}{2005}).

\bibitem[{\citenamefont{Silva and Levit}(2003)}]{Silva01}
\bibinfo{author}{\bibfnamefont{A.}~\bibnamefont{Silva}} \bibnamefont{and}
  \bibinfo{author}{\bibfnamefont{S.}~\bibnamefont{Levit}},
  \bibinfo{journal}{Europhys. Lett.} \textbf{\bibinfo{volume}{62}},
  \bibinfo{pages}{103} (\bibinfo{year}{2003}).

\bibitem[{\citenamefont{Avinun-Kalish et~al.}(2004)\citenamefont{Avinun-Kalish,
  Heiblum, Silva, Mahalu, and Umansky}}]{Avinun01}
\bibinfo{author}{\bibfnamefont{M.}~\bibnamefont{Avinun-Kalish}},
  \bibinfo{author}{\bibfnamefont{M.}~\bibnamefont{Heiblum}},
  \bibinfo{author}{\bibfnamefont{A.}~\bibnamefont{Silva}},
  \bibinfo{author}{\bibfnamefont{D.}~\bibnamefont{Mahalu}} \bibnamefont{and}
  \bibinfo{author}{\bibfnamefont{V.}~\bibnamefont{Umansky}}
  \bibinfo{journal}{Phys. Rev. Lett.} \textbf{\bibinfo{volume}{92}},
  \bibinfo{pages}{156801} (\bibinfo{year}{2004}).

\end{thebibliography}

\end{document}